\documentstyle[tighten,aps,axodraw]{revtex}
\twocolumn

\newcommand{\be}{\begin{equation}}
\newcommand{\ee}{\end{equation}}
\newcommand{\ba}{\begin{eqnarray}}
\newcommand{\ea}{\end{eqnarray}}

\newcommand{\ep}{\epsilon}

\begin{document}

\title{
%
%preprint number:
%
\[ \vspace{-2cm} \]
\noindent\hfill\hbox{\rm  SLAC-PUB-8321, TTP99-51} \vskip 1pt
\noindent\hfill\hbox{\rm hep-ph/9912391} \vskip 10pt
%
% now the title:
%
%\title{
The three-loop relation between the ${\overline {\rm MS}}$ and
the pole quark masses
}

\author{Kirill Melnikov\thanks{
e-mail:  melnikov@slac.stanford.edu}}
\address{Stanford Linear Accelerator Center\\
Stanford University, Stanford, CA 94309}
\author{Timo van Ritbergen\thanks{
e-mail:  timo@particle.physik.uni-karlsruhe.de}}
\address{Institut f\"{u}r Theoretische Teilchenphysik,\\
Universit\"{a}t Karlsruhe,
D--76128 Karlsruhe, Germany}
\maketitle

\begin{abstract}
The analytic relation between the ${\overline {\rm MS}}$
and the pole quark masses is computed to ${\cal O}(\alpha_s^3)$
in QCD. Using this exact result, the accuracy of the large $\beta_0$
approximation is critically examined and the implications of 
the obtained relation for semileptonic $B$ decays are
discussed.
\end{abstract}

%\pacs{??}

\section{Introduction} 

Quark masses, being input parameters of the Standard Model Lagrangian, 
play a very important role in  high energy physics phenomenology. 
It is believed  
%%%The current understanding is 
that the quark masses are generated through
a spontaneous symmetry breaking mechanism caused by a non-zero
vacuum expectation value of a Higgs field. Nevertheless, the quark
masses remain free parameters of the Lagrangian and have to be determined 
  by comparing theoretical predictions with experimental data.

 One has to keep in mind, however, that quarks have not been observed as
 free particles and the notion of a quark mass relies  on
 a theoretical construction.
 Different definitions of quark masses exist referring to different
  schemes for the renormalization of the QCD Lagrangian of the strong 
  interactions.
  Two prominent 
   mass definitions are 
  the $\overline{\rm MS}$  mass $\overline{m}$ and the pole mass $M$.
  The pole mass $M$ is the renormalized quark mass in the on-shell (OS)
  renormalization scheme, while the $\overline{\rm MS}$ mass is the
  renormalized quark mass in the modified minimal subtraction scheme 
 \cite{msscheme,bardeen} which is intimately related to the use of dimensional 
  regularization \cite{thooft}. 

 These two renormalization schemes are used in different physical 
 situations.
 In the $\overline{\rm MS}$ scheme, the renormalized mass and coupling constant
 depend explicitly on the renormalization scale $\mu$ which is often 
  chosen to be of the order of the characteristic scale $Q$ of a process.
  The scale evolution of the $\overline{\rm MS}$
   mass and the coupling constant can be 
  done accurately by integrating the renormalization group equations,
   even for relatively low $Q$.  
  However, 
  the  $\overline{\rm MS}$ mass is by construction
   sensitive to only the short distance (Euclidean) 
  aspects of QCD.
 For processes in which the characteristic 
  scale is large compared to the quark masses it is therefore advantageous
  to adopt the $\overline{\rm MS}$ scheme for the quark mass 
  renormalization\footnote{This is demonstrated for instance in the extraction
  of the ${\overline {\rm MS}}$ $b$-quark mass renormalized at the 
     scale $M_Z$, from three jet events at LEP  with a relatively small 
   uncertainty\cite{delphi}.}.

On the other hand,  for processes where the characteristic scale is set by the 
 mass of a quark in the initial or final state, the situation is different
  as long-distance aspects of QCD become important.
Since in such cases the quarks are  close to the mass 
shell, the on-shell scheme is a natural renormalization scheme
and the pole quark mass emerges.
Explicit perturbative calculations \cite{tarrach,broadhurst,fleischer} 
have shown that the pole quark mass 
is an infrared finite, gauge invariant 
quantity\footnote{This result is proven to all orders in the coupling
constant in Refs. \cite{tarrach,kronfeld}.} and for this reason the pole
mass of a heavy quark has often been considered as a meaningful 
physical parameter with a corresponding numerical value. 

It should be anticipated however that the pole mass of a quark is not 
 a physical quantity in a truly non-perturbative sense since the confiment 
 of quarks in QCD implies that there is no pole in the quark propagator 
  beyond the perturbation theory.
For this reason it is natural to expect that the pole quark mass 
might be sensitive to long distance effects responsible for keeping 
quarks and gluons confined. 
 This expectation was confirmed recently when it became clear 
  that the infrared sensitivity
 of the pole mass reveals itself through contributions in 
 the perturbative series for the pole mass that
 grow factorially at higher orders corresponding to
  a singularity close to the origin in the Borel plane \cite{bigi,braun}.
%% It became clear only recently that the infrared sensitivity 
%% of the pole mass reveals itself through the fact that  
%% the perturbative series using the pole mass receive contributions that
%% grow rapidly at higher orders corresponding to 
%%  a singularity in the Borel plane very close to the origin.
 This renormalon singularity is important for 
  phenomenology since it implies that, indeed,
%  because the fast growing perturbative 
%  contributions mean that one can not 
%  unambiguously determine a 
the  pole quark mass can not be determined with an accuracy better 
  than $\delta M \approx \Lambda_{\rm QCD}$ \cite{bigi,braun}.

 Taken literary, the renormalon analysis applies to asymptotically high
 orders in perturbation theory.
 However, since the location of the leading renormalon 
 pole in the Borel plane can be correctly obtained in the
 so-called large $\beta_0$ approximation and since in the
 lowest non-trivial order of perturbation theory the terms 
 proportional to $\beta_0$ dominate, it was 
 conjectured that the leading renormalon dominates the growth of
 the perturbative coefficients already in low orders for the pole mass.

It is fair to say that this conjecture is an important ingredient 
 in our understanding of theoretical uncertainties for various  
 low-scale physical processes
 such as semileptonic decays of $B$ mesons, determination of the 
$b$ quark mass from $\Upsilon$ mesons spectrum
or the top quark threshold production cross section.
In all these cases eliminating the pole quark mass from theoretical
formulas is an important step in improving the 
 convergence of the perturbative series.
   Nevertheless, it should be kept in mind that, up to date, 
  this physical picture has been checked against only one non-trivial
  order of perturbation theory.

 In spite of its infrared sensitivity, the pole mass remains to be important
 for processes where the on-shell mass definition 
  has technical advantages.
 The pole mass can 
  afterwards be eliminated in favor of a mass that is free from 
   long distance ambiguities such as e.g. the $\overline{\rm MS}$ mass. 
 It often happens that the $\overline{\rm MS}$ mass is not the best 
 choice to parameterize  the low-scale processes and  recently 
 other short-distance  low-scale masses have been 
 proposed\cite{kinetic,1s,ps}. An attractive feature of these
 masses  is that they can be accurately determined from the analysis of 
 the low energy data \cite{hoang,my,beneke}. However, for technical
  reasons, the  relation of these masses to the $\overline{\rm MS}$ mass 
 goes via the pole mass and therefore a determination of the 
 $\overline{\rm MS}$ mass from the low-scale short-distance masses
 requires the knowledge of the conversion formula between 
 $\overline{\rm MS}$ and the pole mass.

  Hence, the relation between 
  the $\overline{\rm MS}$ mass and the pole quark mass
  is  of prime phenomenological importance.
  In the one-loop order this mass relation was obtained in \cite{tarrach}. 
  The first analytical calculation of the two-loop contribution was 
  performed in \cite{broadhurst} and this result was confirmed 
  in \cite{fleischer}. The three-loop coefficient was first
  estimated in Ref.\cite{bbb0} using the large $\beta_0$ approximation. 
  Recently a more accurate result was obtained using asymptotic expansions
  and Pade improvements \cite{chst}.

In this letter we present the analytical calculation of the three-loop
  relation between the $\overline{\rm MS}$ and the pole mass.
Using this exact result, we critically examine the 
validity of the large $\beta_0$ approximation and discuss
some implications of the obtained relation for semileptonic
$B$ decays. A more detailed description of our calculation 
will be published elsewhere.

\section{The three-loop relation between the $\overline {\rm MS}$ and the 
pole masses}

   The quark masses in  both, the $\overline {\rm MS}$ and the on-shell, 
   renormalization schemes are renormalized
   multiplicatively and the connection between renormalized and bare quark 
   masses is defined as 
\ba
   m_{0} && = Z^{\overline{\rm MS}}_m \overline{m},  \\
   m_{0} && = Z^{\rm OS}_m M,  
\ea
where $m_0$ is the bare (unrenormalized) quark mass, and 
  $Z^{\overline{\rm MS}}_m$ and $Z^{\rm OS}_m$ are the renormalization
  factors for the quark mass in, respectively, the 
   $\overline{\rm MS}$ and the on-shell schemes. 
 The relation between the pole quark mass and the $\overline{\rm MS}$ mass
 is then expressed as the ratio
 \ba  \label{Zratio}
   \frac{\overline{m}}{M} = \frac{Z^{\rm OS}_m}{Z^{\overline{\rm MS}}_m}. 
\ea
  One sees that in order to calculate the relation between the 
  $\overline{\rm MS}$ and the pole quark mass 
  in the three-loop order one needs 
  to calculate the mass renormalization factors in both the 
  $\overline{\rm MS}$ and  the on-shell schemes in the three-loop order.
  Fortunately, due to the very simple form of renormalization factors in 
  the $\overline{\rm MS}$ scheme, the mass renormalization factor 
  $Z^{\overline{\rm MS}}_m$ is presently known to a sufficiently high 
  order in perturbation theory and can be taken from the 
  literature \cite{andim3}. One obtains:
\be
  Z^{\overline{\rm MS}}_m = 1 
+\sum \limits_{i=1}^{\infty} C_i \left ( \frac {\alpha_s(\mu)}{\pi} \right )^i,
\label{zms}
\ee
with
\ba
C_1 && = -\frac {1}{\varepsilon}, \nonumber \\
C_2 && = \frac {1}{\varepsilon^2}
         \left ( \frac {15}{8} - \frac {1}{12} N_f \right )
+ \frac {1}{\varepsilon}\left (  - \frac {101}{48} + \frac {5}{72}N_f \right ),
\nonumber \\
C_3 && =
 \frac {1}{\varepsilon^3}\left (  - \frac {65}{16} + \frac {7}{18}N_f 
 - \frac {1}{108}N_f^2 \right )
\nonumber \\
&& + \frac {1}{\varepsilon^2}\left ( \frac {2329}{288} - \frac {25}{36}N_f 
 + \frac {5}{648}N_f^2 \right )
\nonumber \\
&& + \frac {1}{\varepsilon}
      \left (  - \frac {1249}{192} + \frac {5}{18}\zeta_3 N_f 
   + \frac {277}{648}N_f + \frac {35}{3888}N_f^2 \right ),
\ea
where $N_f$ denotes the number of different fermion flavors. The
  dimensional regularization parameter  $\varepsilon$
  is defined through  $\varepsilon = (4-D)/2$ with $D$ being the 
  space-time dimension.

  In comparison to the $\overline{\rm MS}$-scheme, the calculation 
  of renormalization factors in the on-shell scheme is much more involved 
  and for this reason the renormalization factor $Z^{\rm OS}_m$ is 
  known only in the two-loop approximation.
  In the present work we calculate $Z^{\rm OS}_m$ in the three-loop order of
  QCD, thereby obtaining the  $\overline{\rm MS}$ pole mass relation in the
  three-loop order.

Let us explain how the on-shell renormalization factor 
$Z^{\rm OS}_m$ is computed. This renormalization constant follows from
 considering the one particle irreducible quark self-energy parameterized as 
\be
\hat \Sigma(p,M) = M \Sigma_1(p^2,M) + (\hat p - M)\Sigma_2(p^2,M).
\ee
  Here $M$ is the pole-quark-mass and it is understood that the mass
  renormalization is performed in the on-shell scheme which means that 
  on-shell mass counterterms proportional to $M \Sigma_1(M^2,M)$ are 
  inserted in the diagrams of lower order in $\alpha$. 
  For the renormalization of the strong coupling constant we 
  adopt the minimal subtraction scheme
  \be    \frac{\alpha_0}{\pi} =  
                 \frac{\alpha}{\pi}
            - \frac{\beta_0}{\varepsilon}\left( \frac{\alpha}{\pi} \right)^2
            + \left( \frac{\beta_0^2}{\varepsilon^2}
                     -  \frac{\beta_1}{2\varepsilon} \right)
              \left( \frac{\alpha}{\pi} \right)^3   + O(\alpha^4), 
  \ee
 where $\beta_0 = 11/4 - 1/6 N_f $ and $\beta_1 = 51/8 -19/24 N_f$ are the
 first two coefficients of the QCD beta function.
 To make the connection with the formal on-shell mass renormalization factor 
 $Z^{\rm OS}_m$ we note that the perturbative quark propagator can be 
  written as
\be
\hat S_F(p) = \frac {i}{\hat p - m_0 + \hat \Sigma(p,M)}.
\ee
 Since the pole mass of the quark $M$ corresponds to the position
 of the pole of the quark propagator one obtains
  \be  \label{ZOSsigma}
    Z^{\rm OS}_m = 1 + \left. \Sigma_1(p^2,M)\right|_{p^2=M^2} 
 \ee
  which agrees with the form of the on-shell mass counterterms discussed above.

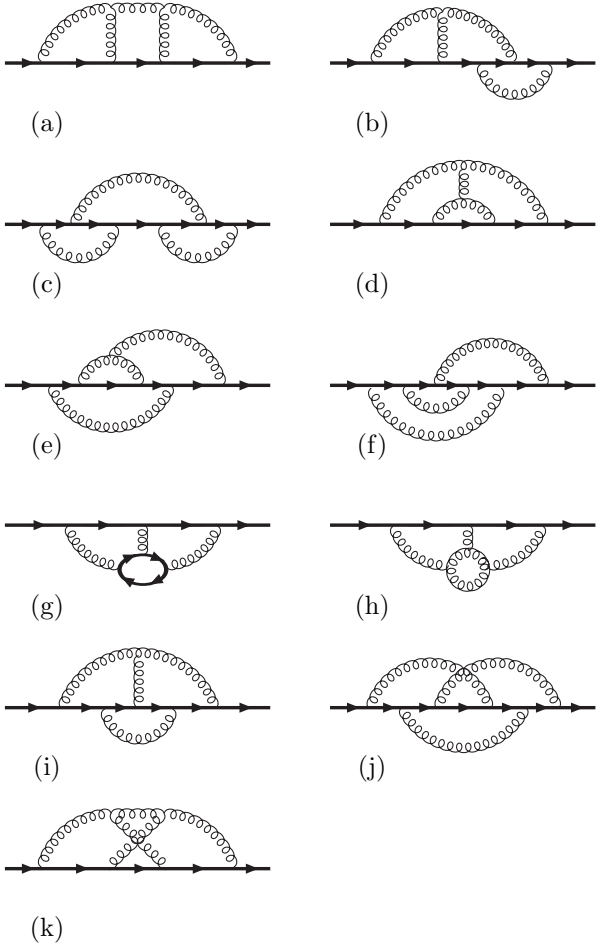
\begin{figure}
\begin{center}
\hfill
\begin{picture}(120,40)(0,0)
  \SetScale{.8}
 \SetWidth{1.6}
 \Line(5,20)(130,20)
 \SetWidth{1}
  \ArrowLine(10,20)(22,20)
 \ArrowLine(35,20)(48,20)
 \ArrowLine(60,20)(80,20) 
 \ArrowLine(83,20)(110,20) 
 \ArrowLine(118,20)(125,20)
 \SetWidth{0.5}
 \GlueArc(80,12)(34,15,90){2.2}{9.5}  
 \GlueArc(55,12)(34,90,165){2.2}{9.5}
 \Gluon(80,20)(80,46){2.2}{5}
 \Gluon(55,20)(55,46){-2.2}{5}
 \Gluon(80,46)(55,46){-2.2}{4}
 \Text(20,-2)[t]{(a)}
\end{picture}
\hfill
\begin{picture}(120,40)(0,0)
 \SetScale{.8}
 \SetWidth{1.6}
 \Line(5,20)(130,20)
 \SetWidth{1}
  \ArrowLine(10,20)(22,20)
 \ArrowLine(38,20)(48,20)
 \ArrowLine(60,20)(80,20)
 \ArrowLine(82,20)(90,20)
 \ArrowLine(92,20)(110,20)
 \ArrowLine(114,20)(125,20)
 \SetWidth{0.5}
 \GlueArc(58,8)(36,21,90){2.2}{10.5}
 \GlueArc(58,10)(34,90,162){2.2}{9.5}
 \Gluon(58,20)(58,44){2.2}{5}
 \GlueArc(92,21)(16,186,354){2.2}{9.5}
\Text(20,-2)[t]{(b)}
\end{picture}
\hfill\null\\
\hfill
\begin{picture}(120,60)(0,0)
 \SetScale{.8}
 \SetWidth{1.6}
 \Line(5,20)(130,20)
 \SetWidth{1}
 \ArrowLine(10,20)(22,20)
 \ArrowLine(24,20)(40,20)
 \ArrowLine(40,20)(55,20)
 \ArrowLine(60,20)(80,20)
 \ArrowLine(86,20)(96,20)
 \ArrowLine(104,20)(110,20)
 \ArrowLine(119,20)(125,20)
 \SetWidth{0.5}
 \GlueArc(96,21)(17,186,354){2.2}{9.5}
 \GlueArc(40,21)(17,186,354){2.2}{9.5}
 \GlueArc(68,10)(32,17,163){2.2}{17.5}
 \Text(20,-2)[t]{(c)}
\end{picture}
\hfill
\begin{picture}(120,60)(0,0)
  \SetScale{.8} 
 \SetWidth{1.6}
 \Line(5,20)(130,20)
 \SetWidth{1}
  \ArrowLine(16,20)(22,20)
 \ArrowLine(36,20)(50,20)
 \ArrowLine(60,20)(80,20)
 \ArrowLine(90,20)(100,20)  
 \ArrowLine(112,20)(125,20)
 \SetWidth{0.5}
 \GlueArc(68,16)(14,17,163){2.2}{7.5}
 \GlueArc(68,8)(40,17,163){2.2}{21.5}
  \Gluon(68,32)(68,46){2.2}{3.5}
 \Text(20,-2)[t]{(d)}
\SetScale{1}
\end{picture}
\hfill\null\\
\hfill
\begin{picture}(120,60)(0,0)
 \SetScale{.8}
 \SetWidth{1.6}
 \Line(5,20)(130,20)
 \SetWidth{1}
  \ArrowLine(10,20)(22,20)
 \ArrowLine(28,20)(42,20)
 \ArrowLine(50,20)(62,20)
 \ArrowLine(68,20)(87,20)
 \ArrowLine(94,20)(99,20)  
 \ArrowLine(112,20)(125,20)
 \SetWidth{0.5}
 \GlueArc(55,18)(14,10,170){2.2}{8.5}   
 \GlueArc(55,30)(30,200,340){2.2}{16.5}   
 \GlueArc(78,14)(30,12,140.5){2.2}{14}  
 \Text(20,-2)[t]{(e)}
\end{picture}
\hfill
\begin{picture}(120,60)(0,0)
  \SetScale{.8}
 \SetWidth{1.6}
 \Line(5,20)(130,20)
 \SetWidth{1}
  \ArrowLine(10,20)(22,20)
 \ArrowLine(25,20)(42,20)
 \ArrowLine(42,20)(58,20)  
 \ArrowLine(58,20)(68,20)
 \ArrowLine(68,20)(87,20)
 \ArrowLine(92,20)(100,20)
 \ArrowLine(112,20)(125,20)
 \SetWidth{0.5}
 \GlueArc(55,24)(15,197,343){2.2}{7.5}
 \GlueArc(55,28)(32,197,343){2.2}{16.5}
 \GlueArc(81,14)(26,13,167){2.2}{16.5}
 \Text(20,-2)[t]{(f)}
\end{picture}
\hfill\null\\
\hfill
\begin{picture}(120,60)(0,0)
    \SetScale{.8}
 \SetWidth{1.6}
 \Line(5,30)(130,30)
  \SetWidth{1.3}
  \Oval(70,9)(7,11)(0)
 \SetWidth{1}
 \ArrowLine(15,30)(28,30)
 \ArrowLine(42,30)(62,30)
 \ArrowLine(83,30)(99,30)
 \ArrowLine(115,30)(120,30)
  \ArrowLine(61.5,14)(66.5,16)
  \ArrowLine(74,15.5)(80,13)
   \ArrowLine(79,5.0)(73,2)
   \ArrowLine(65,2.7)(60,5.5)
 \SetWidth{0.5}
  \Gluon(70,30)(70,16){2.2}{3} 
 \GlueArc(59,36)(25,195,270){2.2}{7.5}
 \GlueArc(81,36)(25,270,345){2.2}{7.5}
 \Text(20,-2)[t]{(g)}
\end{picture}
\hfill
\begin{picture}(120,60)(0,0)
 \SetScale{.8}
 \SetWidth{1.6}
 \Line(5,30)(130,30)
  \SetWidth{1.3}
 \SetWidth{1}
 \ArrowLine(15,30)(28,30)
 \ArrowLine(42,30)(62,30)  
 \ArrowLine(83,30)(99,30)
 \ArrowLine(115,30)(120,30) 
 \SetWidth{0.5}
 \GlueArc(70,9)(8,0,400){2.2}{12.8}
 \Gluon(70,30)(70,18){2.2}{2.5}
 \GlueArc(59,36)(25,195,274){2.2}{7.5}
 \GlueArc(81,36)(25,266,345){2.2}{7.5}
 \Text(20,-2)[t]{(h)}
\end{picture}
\hfill\null\\
\hfill
\begin{picture}(120,60)(0,0)
 \SetScale{.8}
 \SetWidth{1.6}
 \Line(5,20)(130,20)
 \SetWidth{1}
  \ArrowLine(16,20)(22,20)
 \ArrowLine(38,20)(50,20)
 \ArrowLine(50,20)(70,20)
 \ArrowLine(74,20)(80,20)
 \ArrowLine(92,20)(100,20)
 \ArrowLine(112,20)(125,20)
 \SetWidth{0.5}
 \GlueArc(68,8)(38,18,162){2.2}{21.5}
  \Gluon(68,20)(68,44.5){2.2}{5.5}
  \GlueArc(68,21)(16,186,354){2.2}{9.5}  
 \Text(20,-2)[t]{(i)}
\end{picture}
\hfill
\begin{picture}(120,60)(0,0)
 \SetScale{.8} 
 \SetWidth{1.6}
 \Line(5,20)(130,20)
 \SetWidth{1}
 \ArrowLine(10,20)(22,20)
 \ArrowLine(24,20)(42,20) 
 \ArrowLine(42,20)(58,20) 
 \ArrowLine(59,20)(78,20)
 \ArrowLine(82,20)(96,20) 
 \ArrowLine(100,20)(110,20)
 \ArrowLine(118,20)(125,20)
 \SetWidth{0.5}
 \GlueArc(84,12)(29.2,15,165){2.2}{16.5}
 \GlueArc(68,33)(32,204,336){2.2}{16.5}
 \GlueArc(52,12)(29.2,15,165){2.2}{16.5}
\Text(20,-2)[t]{(j)}
\end{picture}
\hfill\null\\
\hfill
\begin{picture}(120,60)(0,0)
 \SetScale{.8}
\SetWidth{1.6}
 \Line(5,20)(130,20)
 \SetWidth{1}
  \ArrowLine(10,20)(22,20)
 \ArrowLine(35,20)(48,20)
 \ArrowLine(58,20)(80,20)
 \ArrowLine(83,20)(110,20)
 \ArrowLine(118,20)(125,20)
 \SetWidth{0.5}
 \GlueArc(80,12)(34,15,90){2.2}{9.5}
 \GlueArc(55,12)(34,90,165){2.2}{9.5}
  \Gluon(82,20)(55,46){2.2}{7.5}
  \Gluon(53,20)(80,46){-2.2}{7.5}
 \Gluon(80,46)(55,46){-2.2}{4}
 \Text(20,-2)[t]{(k)}
\end{picture}
\hfill
\begin{picture}(120,60)(0,0)
\end{picture}
\hfill\null\\
\vglue 18pt
\end{center}
\caption{Examples of three-loop quark propagator diagrams corresponding to
   eleven integration topologies.}
\label{fig:examplediagrams}
\end{figure}

 Equation (\ref{ZOSsigma}) provides the simplest formula for the calculation
 of $Z^{\rm OS}_m$. One computes $\Sigma_1(M^2,M)$ to the required order 
  in perturbation theory taking into account the lower order difference 
  between $m_0$ and $M$ by calculating lower order diagrams with the 
  appropriate mass counterterm insertions.
Therefore, in order to obtain the three-loop on-shell
renormalization factor $Z^{\rm OS}_m$ one has to compute 
  on-shell propagator-type diagrams up to three loops.

The most efficient way to evaluate those multiloop integrals 
is to utilize integration-by-parts identities within dimensional 
regularization. There are eleven basic topologies that should be 
considered\footnote{For Abelian gauge theory, like QED, 
the number of basic topologies would be just four.}. They are shown in 
Fig.1.  For each of the topologies one writes down a system of 
recurrence relations based on integration-by-parts identities 
\cite{thooft,ibp}.
Solving this system, it is possible to show that any integral which 
belongs to the above topologies can be expressed through $18$ 
master integrals. Fortunately, most of these integrals have  been 
computed in the course of the analytical calculation 
of the electron anomalous magnetic moment 
\cite{Laporta} and can be taken from there.
As compared to that reference, we need one additional
master integral and we also need one of the master integrals
of Ref. \cite{Laporta} to a higher order in the regularization
parameter $\varepsilon$.
For completeness, we present here the results for these two master 
integrals.  Let us introduce
$D_1=k_1^2, D_2=k_2^2,D_3=k_3^2,D_4=(k_1-k_2)^2,D_5=(k_3-k_2)^2,
D_6 = k_1^2 + 2pk_1, D_7 = k_2^2 + 2pk_2, D_8= k_3^2 + 2pk_3$
with $p^2=-1$.
Then:
$$
I_{18} = 
 \int \frac {{\rm d}^Dk_1 {\rm d}^Dk_2  {\rm d}^Dk_3}
 {D_1 D_3 D_4 D_5 D_6 D_7 D_8} = C(\varepsilon) \left (2 \pi^2 \zeta_3 
 - 5 \zeta_5 \right ),
$$
where $C(\varepsilon)=\left [ \pi^{(2-\varepsilon)} \Gamma(1+\varepsilon) 
     \right ]^3$ and
  $\zeta_k= \zeta(k)$ denotes the Riemann zeta function.
We also give the result for $I_{10}$ (we use notations as in \cite{Laporta})
 valid to ${\cal O}(\varepsilon^2)$:
\ba
&& I_{10} = \int \frac {{\rm d}^Dk_1 {\rm d}^Dk_2  {\rm d}^Dk_3}
 {D_2 D_4 D_5 D_6  D_8} = C(\varepsilon)
 \left \{ \frac {-1}{3\varepsilon^3} -\frac {5}{3\varepsilon^2}
\right. \nonumber \\
&& \left. + \frac {1}{\varepsilon} \left ( -\frac {2\pi^2}{3} -4  \right )
 -\frac {26}{3}\zeta_3 - \frac {7\pi^2}{3}+\frac {10}{3}
\right. \nonumber \\
&& \left.
+ \varepsilon \left ( -\frac {35\pi^4}{18} -\frac {94}{3}\zeta_3
 -\pi^2 + \frac {302}{3} \right ) 
  + \varepsilon^2 \left (  734 - \frac {76\pi^2}{3} \zeta_3
\right. \right.
\nonumber \\
&& \left. \left.
 + 
   \frac {101 \pi^2}{3} - 20\zeta_3 - \frac {551}{90}\pi^4 
 - 462\zeta_5 \right ) \right \}. 
\nonumber
\ea

There are two principal checks on our solution of the recurrence relations.
First, we have computed the three-loop anomalous magnetic moment of the 
 electron and confirmed the result of Ref.\cite{Laporta}. Second, the actual
calculation of the on-shell  quark mass renormalization constant 
$Z^{\rm OS}_m$ has been performed in  an arbitrary covariant gauge 
for the gluon field. The explicit
cancellation of the gauge parameter in our result for $Z^{\rm OS}_m$
is an important check of the correctness of the calculation.

Having computed the three-loop contribution to $Z^{\rm OS}_m$ in this way 
and using  Eqs.(\ref{Zratio},\ref{zms}) 
we obtain the three-loop relation between the $\overline{\rm MS}$ quark
mass and the pole mass.
Below we present the exact result only for this mass relation but we should
note for future reference that $Z^{\rm OS}_m$ is easily recovered
using Eq.(\ref{Zratio}).
The relation between the masses is expressed in terms of color 
 and flavor factors as
\ba
{\overline m}(M) && = M \left[ 
1 - C_F \left ( \frac {\alpha_s}{\pi} \right )
+ C_F \left ( \frac {\alpha_s}{\pi} \right )^2 
\left (C_F\, d^{(2)}_1 +  
\right. \right. \nonumber \\
&& \left. \left.
 + C_A\, d^{(2)}_2 
+ T_RN_L\,d^{(2)}_3
+T_RN_H \, d^{(2)}_4
\right )
\right. \nonumber \\
&& \left.
+C_F  \left ( \frac {\alpha_s}{\pi} \right )^3
 \left ( 
 C_F^2\, d^{(3)}_1
 + C_FC_A \, d^{(3)}_2 
 + C_A^2\, d^{(3)}_3
\right. \right. \nonumber \\
&& \left. \left.
 + C_F T_R N_L \, d^{(3)}_4
 + C_F T_R N_H \, d^{(3)}_5
 + C_A T_R N_L \, d^{(3)}_6
\right. \right. \nonumber \vspace{4mm} \\ 
&& \left. \left.
 + C_A T_R N_H \, d^{(3)}_7
 + T_R^2 N_L N_H \, d^{(3)}_8
 + T_R^2 N_H^2 \, d^{(3)}_{9}
\right. \right. \nonumber \\
&& \left. \left.
 + T_R^2 N_L^2 \, d^{(3)}_{10}
\right )   + {\cal O}\left( \frac {\alpha_s}{\pi} \right)^4  \right],
 \label{mainresult}
\ea
where $C_F$ and $C_A$ are the Casimir operators of the fundamental and the
adjoint representation of the color gauge group (the group is SU(3) 
 for QCD) and $T_R$ is the trace normalization of the 
  fundamental representation.
  $N_L$ is the number of massless quark flavors and $N_H$ is the number of 
  quark flavors with a pole mass equal to $M$. Also
 $\alpha_s \equiv \alpha_s^{(N_L+N_H)}(M)$ is the ${\overline {\rm MS}}$ 
 strong coupling constant renormalized at the scale of the pole mass $\mu = M$
 in the theory with $N_L+N_H$ active flavors\footnote{
 We should note that, if needed, Eq.(\ref{mainresult}) is easily expressed 
 in terms of $\overline {m}(\mu)$, $\alpha_s(\mu)$ and $\log(\mu/M)$
 by using  the renormalization group equations for $\overline {m}(\mu)$ 
 and $\alpha_s(\mu)$.}. 

Our calculation yields the following result for the coefficients 
 $d^{(n)}_k$ in Eq. (\ref{mainresult}):
\ba
d^{(2)}_1 && =  \frac {7}{128} - \frac {3}{4}\zeta_3 
 + \frac {1}{2}\pi^2 \log2 - \frac {5}{16}\pi^2, 
\nonumber \\
d^{(2)}_2 && =   
 - \frac {1111}{384} + \frac {3}{8}\zeta_3 - \frac {1}{4}\pi^2 \log2 
 + \frac {1}{12} \pi^2, 
\nonumber \\
d^{(2)}_3 && =  \frac {71}{96} + \frac {1}{12}\pi^2, 
\nonumber \\
 d^{(2)}_4 && =  \frac {143}{96} - \frac {1}{6}\pi^2, 
\nonumber \\
 d^{(3)}_1 && =
 -\frac {2969}{768} -\frac {1}{16}\pi^2 \zeta_3 
 - \frac {81}{16}\zeta_3 + \frac {5}{8}\zeta_5
 + \frac {29}{4}\pi^2\log2 
 \nonumber \\
&&
 + \frac {1}{2} \pi^2 \log^22 
 - \frac {613}{192}\pi^2 - \frac {1}{48}\pi^4 
 - \frac {1}{2}\log^4 2  - 12a_4, 
\nonumber \\
 d^{(3)}_2 && =
 \frac {13189}{4608} - \frac {19}{16}\pi^2 \zeta_3 - \frac {773}{96}\zeta_3 
+ \frac {45}{16}\zeta_5 - \frac {31}{72}\pi^2 \log 2 
 \nonumber \\
&&   
 - \frac {31}{36}\pi^2 \log^2 2 
 + \frac {509}{576} \pi^2+ \frac {65}{432} \pi^4
 - \frac {1}{18}\log^4 2  - \frac {4}{3}a_4, 
\nonumber \\
 d^{(3)}_3 && =
 -\frac{1322545}{124416} + \frac{51}{64}\pi^2 \zeta_3 + 
\frac {1343}{288}\zeta_3 - \frac {65}{32}\zeta_5
 \nonumber \\ 
&&  
 -\frac {115}{72}\pi^2\log 2 + \frac {11}{36}\pi^2 \log^2 2 
 -\frac {1955}{3456}\pi^2
          - \frac {179}{3456}\pi^4 
 \nonumber \\
&&  + \frac {11}{72}\log^4 2 
 + \frac {11}{3}a_4, 
\nonumber \\
 d^{(3)}_4 && = 
 \frac {1283}{576} + \frac {55}{24}\zeta_3 
 - \frac {11}{9}\pi^2 \log 2 + \frac {2}{9}\pi^2 \log^2 2 
 + \frac {13}{18} \pi^2
 \nonumber \\
&&  
 - \frac {119}{2160}\pi^4 
 + \frac {1}{9}\log^4 2 + \frac {8}{3}a_4, 
\nonumber \\
 d^{(3)}_5 && = 
\frac {1067}{576} - \frac {53}{24}\zeta_3 + \frac {8}{9}\pi^2 \log 2 
 - \frac {1}{9}\pi^2 \log^2 2 
 - \frac {85}{108}\pi^2 
 \nonumber \\
&& 
+ \frac {91}{2160}\pi^4 
 + \frac {1}{9}\log^4 2 + \frac {8}{3}a_4, 
\nonumber \\
 d^{(3)}_6 && = 
 \frac{70763}{15552} + \frac {89}{144}\zeta_3 
 + \frac {11}{18}\pi^2\log 2
 - \frac {1}{9} \pi^2 \log^2 2  
 \nonumber \\
&&  + \frac {175}{432}\pi^2  
 + \frac {19}{2160}\pi^4 
 -\frac {1}{18}\log^4 2 - \frac {4}{3}a_4, 
\nonumber \\
 d^{(3)}_7 && = 
\frac {144959}{15552} + \frac {1}{8} \pi^2 \zeta_3 
 - \frac {109}{144}\zeta_3
 - \frac {5}{8}\zeta_5 
 + \frac {32}{9}\pi^2 \log 2 
 \nonumber \\
&&
+ \frac {1}{18} \pi^2 \log^2 2 
 - \frac {449}{144}\pi^2 - \frac {43}{1080}\pi^4 
 - \frac {1}{18}\log^4 2 - \frac {4}{3}a_4, 
\nonumber \\
d^{(3)}_8 && = 
  - \frac {5917}{3888} + \frac {2}{9}\zeta_3 + 
 \frac {13}{108}\pi^2, 
\nonumber \\
 d^{(3)}_{9} && = 
 -\frac {9481}{7776} + \frac {11}{18}\zeta_3 + \frac {4}{135}\pi^2,  
\nonumber \\
 d^{(3)}_{10} && = 
  - \frac {2353}{7776} - \frac {7}{18}\zeta_3 - \frac {13}{108}\pi^2. 
\nonumber
\ea
where $a_4 = \sum \limits_{n=1}^{\infty} 1/(2^n n^4) 
   = {\rm Li}_4(1/2) \approx 0.517479$. 

For standard values of the QCD color factors
%\be 
$C_F=4/3,~C_A=3,~~T_R=1/2$,
%\ee
and assuming the number of heavy flavors $N_H$ to be equal to one,
the result reads:
\ba
&&\frac {{\overline m}(M)}{M} = 1
       - \frac {4}{3} \left (\frac {\alpha_s}{\pi} \right )
        + \left (\frac {\alpha_s}{\pi} \right )^2 \left (
       N_L \left ( \frac {71}{144} + \frac {\pi^2}{18} \right )
\right. 
\nonumber \\
&& \left.       - \frac {3019}{288} + \frac {1}{6}\zeta_3 
         - \frac {\pi^2}{9}\log 2 - \frac {\pi^2}{3}
 \right )
\nonumber \\
&& + \left (\frac {\alpha_s}{\pi} \right )^3 \left (
        N_L^2  \left (  - \frac {2353}{23328} - \frac {7}{54}\zeta_3 
         - \frac {13}{324}\pi^2 \right )
\right.
\nonumber \\
&& \left.      
        + N_L  \left ( \frac {246643}{23328} + \frac {241}{72}\zeta_3 + 
         \frac {11}{81}\pi^2\log 2 
         - \frac {2}{81} \pi^2 \log^2 2 
\right. \right.
\nonumber \\
&& \left. \left.
   + \frac {967}{648}\pi^2 - \frac {61}{1944}\pi^4 
          - \frac {1}{81}\log^4 2  - \frac {8}{27}a_4 
          \right )
\right.
\nonumber \\
&& \left. 
         - \frac {9478333}{93312} + \frac {1439}{432}\zeta_3\pi^2 
          - \frac {61}{27}\zeta_3 - 
         \frac {1975}{216}\zeta_5 
\right.
\nonumber \\
&& \left. 
          + \frac {587}{162}\pi^2 \log 2 
          + \frac {22}{81} \pi^2 \log^2 2 - \frac {644201}{38880}\pi^2
          + \frac {695}{7776} \pi^4
\right.
\nonumber \\
&& \left.
          + \frac {55}{162}\log^4 2 
          + \frac {220}{27} a_4 
\right ).
\label{main}
\ea

Numerically, we obtain:
\ba
&& \frac {{\overline m}(M)}{M} =  
1 - \frac{4}{3} \left ( \frac {\alpha_s}{\pi} \right )
+\left ( \frac {\alpha_s}{\pi} \right )^2
\left ( 1.0414~N_L  -14.3323 \right )
\nonumber \\
&& 
+\left ( \frac {\alpha_s}{\pi} \right )^3
\left ( 
 -0.65269~N_L^2 +26.9239~N_L -198.7068
\right ) .
\label{msthrpole}
\ea

 We note that the exact three-loop  ${\cal O}(N_L)$ 
 and ${\cal O}(N_L^0)$ coefficients are within the
 errors of the values $27.3(7)N_L$ and $202(5)$ obtained in \cite{chst},
 nevertheless for the individual color structures $ C_F C_A^2 $ 
 and $C_F^2 N_L $ the difference between the exact result  
 and the numerical values in \cite{chst} is larger than the error
 bars quoted in that reference.
  
 From Eq.(\ref{main}) a relation
between the pole mass $M$ and the $\overline {m}(\overline {m})$, the 
$\overline {\rm MS}$ mass normalized at the scale
$\mu = \overline {m}(\mu)$ is easily derived. One obtains
% For the renormalization group solution of $\overline {m}(\mu)$ 
% in terms of the invariant mass that is compatible with the $\alpha_s^3$ 
% order in the mass relation results of \cite{4looprenorm} are needed.}:
%
\ba
&&\frac {M}{\overline {m}(\overline {m})} =  
1 + \frac{4}{3} \left ( \frac {\bar \alpha_s}{\pi} \right )
+\left ( \frac {\bar \alpha_s}{\pi} \right )^2
\left ( - 1.0414~N_L  + 13.4434 \right )
\nonumber \\
&&  \label{resultmm}
+\left ( \frac {\bar \alpha_s}{\pi} \right )^3
\left ( 
 0.6527 ~N_L^2
 -26.655 ~N_L
+190.595 
\right ),
\label{polethrms}
\ea
with $\bar \alpha_s \equiv \alpha_s(\overline {m})$.

It is interesting to look at these results when they are parameterized in 
a way, that separates  the effects 
  related to the running of the coupling constant 
  (i.e. contributions that are related to a change of the scale $\mu$ in the 
    coupling constant)
% caused by the running of the
% that can be absorbed in the scale $\mu$ of the
 and the remaining part.
% \footnote{This remaining part
%  would correspond to ``QCD with zero $\beta$-function" and
%  is therfore refered to as the .  
 To do this in a consistent 
manner, we first use the decoupling relations to reexpress
the $\overline {\rm MS}$ coupling constant $\alpha_s$ defined
in the theory with $N_L$ massless $+1$ massive flavors through the 
coupling, defined in the theory with $N_L$ massless quarks
only \cite{dcpl1,dcpl2}. Using the first two
coefficients of the $\beta$-function:
$$
\beta_0 = \frac{1}{4} \left ( 11 - \frac {2}{3} N_L \right ),~~~
\beta_1 = \frac {1}{16} \left ( 102 - \frac {38}{3} N_L \right ),
$$
we obtain:
\ba
&& \frac {M}{\overline{m}(\overline{m})} =  
1 + \frac{4}{3} \left ( \frac {\bar \alpha_s}{\pi} \right )
+\left ( \frac {\bar \alpha_s}{\pi} \right )^2
\left ( 6.248~\beta_0 -3.739 \right )
\nonumber \\
&& 
+\left ( \frac {\bar \alpha_s}{\pi} \right )^3
\left ( 
 23.497~\beta_0^2
 + 6.248~\beta_1
 + 1.019~\beta_0
 -29.94
\right ),
\label{conf}
\ea
where $\bar \alpha_s$ now stands for the $\overline {\rm MS}$ coupling 
defined in the theory with $N_L$ fermions only.

 From the above equation one sees that the 
large $\beta_0$ approximation works well at ${\cal O}(\alpha_s^3)$
especially for $N_L =3$, but in part because of a 
%significant 
cancellation between the conformal (i.e. $\beta$-independent) term
and the term proportional to $\beta_1$. The magnitude of conformal 
terms is large  and they exhibit a very fast 
growth from one  order of perturbation theory to the other. 

In order to check that the magnitude of  conformal terms is not an 
artifact of our use of  the $\overline {\rm MS}$ coupling
constant as an expansion parameter, one can rewrite the  equation for the
mass ratio  in  terms of the  $V$-scheme
coupling $\bar \alpha_V = \alpha_V(\overline {m})$. 
To the necessary order the relation between $\alpha_V$ and $\alpha_s$
is given in Ref.{\cite{schroeder}. Rewriting Eq.(\ref{conf})
in $\alpha_V$, one obtains  the result
which does not improve as compared to the $\overline {\rm MS}$ 
scheme. The overall magnitude of corrections gets somewhat decreased,
but conformal coefficients remain very large and grow rapidly.

This situation makes it difficult to interpret the 
high accuracy of the large $\beta_0$ approximation at 
${\cal O}(\alpha_s^3)$ 
%of Eq.(\ref{conf})
%just 
simply from the assumption of a leading infrared renormalon dominance\footnote{
 We  note that the large $\beta_0$ limit and the 
infrared renormalon dominance are not the same thing.
Also in the renormalon context corrections to the large $\beta_0$ limit 
 grow with the order in perturbation theory and eventually must
  make the large $\beta_0$
limit unreliable. We thank M. Beneke for correspondence on this point.}.
On the one hand we clearly 
see that the large $\beta_0$ approximation remains extremely
successful in the third order of the perturbative expansion for the
  pole mass. 
On the other hand, 
 part of the  reason for this success seems to be a
 cancellation between subleading $\beta$-dependent
 and fast growing conformal terms. Since a different physics is usually
 associated with these contributions 
  it is unclear how this situation will extrapolate
  to the next order in the $\overline{\rm MS}$ pole mass relation. 
  We refrain from drawing a definite conclusion on this point at the moment.

Since exact calculations in even higher orders of perturbation 
theory seem formidably complicated, it is perhaps reasonable 
to look for other approximation schemes which, as compared to the large 
$\beta_0$ approximation, will preserve in a better way the field 
theoretical structure of QCD.  One possibility is the large 
$N_c$ approximation \cite{largenexp}. Its attractive feature
is that only planar diagrams contribute to leading order in $N_c$
and this significantly reduces the technical burden of the calculations.
Since in realistic QCD we are interested in the applications
where $N_c \sim N_L$,  one can go a step further and 
consider a situation where both the number of colors $N_c$ and the number of 
light fermions $N_L$ are considered to be  large
numbers.
The correction to this approximation is expected to
be of the order ${\cal O}(N_c^{-2})$ which for $N_c = 3$ 
%means that the corrections are about $10$ percent
 is about $10$ percent\footnote{For contributions 
 of massive fermion loops the accuracy of the approximation is 
 formally ${\cal O}(N_c^{-1})$ but due to the mass of these fermions 
 these contributions are usually small.}.
The result for the $\overline {\rm MS}$ mass 
in this approximation reads:
\ba
&& \left[ \frac{{\overline m}(M)}{M} \right]_{N_c} = \left ( 
1 - 1.5 \left ( \frac {\alpha_s}{\pi} \right )
+\left ( \frac {\alpha_s}{\pi} \right )^2
\left ( 1.17~N_L  -16.13 \right )
\right. \nonumber \\
&& \left.
+\left ( \frac {\alpha_s}{\pi} \right )^3
\left ( 
 -0.73~N_L^2
 +30.41~N_L
 -222.84
\right ) \right ).
\label{largen}
\ea
Comparing this equation with Eq.~(\ref{msthrpole}) one sees that the 
large $N_c$ approximation works with the expected accuracy and 
hence might serve as a useful starting point for 
the analysis at an even higher order.

\section{The three-loop analysis of semileptonic $B$ decays}

Let us now turn to the analysis of the semileptonic decay
process $B \to X_u e \nu_e$. 
 We want to investigate whether various existing estimates of  
 the ${\cal O}(\alpha_s^3)$ correction to 
 this process are compatible with the notion 
 of an improved quality of the perturbative series 
 at low orders when the pole quark mass is eliminated in favor 
 of a properly defined low-scale short-distance mass.
 Some of these estimates refer explicitly to the $\overline{\rm MS}$ mass
 and this investigation requires the use of the three-loop quark mass relation
  Eq. (\ref{resultmm}).

Recall, that 
the Heavy Quark Expansion demonstrates that 
the non-perturbative effects 
are small
 % ($\approx ??? \%$)
in this process \cite{npb}
%aresuppressed by at least two powers of the heavy b quark mass \cite{npb}
and an important issue is the accurate calculation of the perturbative
corrections to the quark level decay $b \to u e \nu_e$.
To ${\cal O}(\alpha_s^2)$ such a calculation
has been performed in Ref.\cite{timo}. Recently the three-loop 
correction to that process has been estimated in Ref.\cite{BslPade}
using asymptotic Pade approximants. It is unclear to us
at the moment how accurate that estimate is, but we will take
the number quoted in Ref.\cite{BslPade} as a reference point. 
For five active flavors we  write:
\ba
\Gamma_{\rm sl}&& = \frac {G_F^2|V_{ub}|^2 
{\overline {m}(\overline {m})}^5}{192\pi^3} 
\left \{ 1 + 4.25 \left ( \frac {\bar \alpha_s^{(5)}}{\pi}\right )
+ 26.78 \left ( \frac {\bar \alpha_s^{(5)}}{\pi}\right )^2
\right. \nonumber \\
&& \left. 
+ (200 + \Delta ) \left ( \frac {\bar \alpha_s^{(5)}}{\pi}\right )^3
\right \}.
\label{rate0}
\ea
The central value of the third order coefficient is due to 
Ref.\cite{BslPade} and $\Delta$ parameterizes possible 
uncertainties in this estimate. 
For the chosen normalization point $\mu = \overline {m}$, one sees
a rapid increase in the coefficients of the series in Eq.(\ref{rate0}). 
The reason for that  is thought to be a poor choice of the renormalization 
scale, since  a typical energy release in $B$ decays is of the 
order $m_b/5$ rather than $m_b$.  Since at very low scales the logarithmic
running of the quark mass
% , like is the case for the $\overline {\rm MS}$ mass,
is considered unphysical, it was suggested to use other
short-distance masses which have a more transparent meaning at a low
normalization point.
%  An introduction of such masses does help at the two-loop level. 
In this paper we would like to 
check the behavior of perturbation series once the decay width is 
expressed in terms of the so-called $1S$ mass. Strictly speaking,
the $1S$ mass introduced in Ref. \cite{1s} is not a truly 
short-distance quantity and its usefulness for $B$ decays when 
both perturbative and non-perturbative corrections are 
taken into account remains to be proven.
 Nevertheless, we use it here because its 
relation to the pole mass is known to a sufficiently high order 
in the perturbative expansion. It was suggested that the correct
way to incorporate the $1S$ mass into semileptonic $B$ decays
is the $\Upsilon$ expansion \cite{upexp} and we will denote 
different terms in this expansion by a formal parameter $\epsilon$
 (which is eventually put to one). 
We use the relation between the $\overline {\rm MS}$ mass and the pole mass 
to rewrite Eq.(\ref{rate0}) in terms of the 
%pole mass and the 
 $1S$ mass (via its relation to the pole mass) and check
if the result is consistent with 
%various approximations. 
an improved behavior of the perturbation series.
Below we consider four possible scenarios. 

i) The large $\beta_0$ estimate of the three-loop coefficient
of $\Gamma_{\rm sl}$ is  accurate\footnote{When we refer to the
results of ``the large $\beta_0$ approximation'' in this context, 
we have in mind the expression for $\Gamma_{\rm sl}$ written
in terms of the pole mass. The predictions for $\Gamma_{\rm sl}$
in the large $\beta_0$ approximation can be found in Ref.\cite{bbb}.}.  
Using the mass relation,
derived in this paper, we find that this requires 
$\Delta \approx -100$ in Eq.(\ref{rate0}). We then rewrite 
the decay rate in terms of the $1S$ mass and obtain
for $\bar \alpha_s^{(4)} = 0.22$:
$$
\Gamma_{\rm sl} = \frac {G_F^2|V_{ub}|^2 {m_{1S}}^5}{192\pi^3} 
\left ( 
1-0.115\ep-0.031\ep^2-0.027\ep^3
\right ).
$$

ii) Let us assume that the pattern of the $n$-th loop coefficient in 
 Eq.(\ref{rate0}) is well approximated by $5^n$ (see Ref.\cite{bes}). 
For the three-loop coefficient in Eq.(\ref{rate0}) this
implies $\Delta \approx -75$. 
 Expressed in terms of the pole mass the three-loop coefficient of the
 width is then a factor 0.9 smaller than the large $\beta_0$ estimate.
%In this case the large $\beta_0$ estimate
%is about $10$ per cent above
% \footnote{We mean here the absolute value.} 
%the true value. 
Expressed through the
$1S$ mass, the decay width for $\bar \alpha_s^{(4)} = 0.22$ reads:
$$
\Gamma_{\rm sl} = \frac {G_F^2|V_{ub}|^2 {m_{1S}}^5}{192\pi^3} 
\left ( 
1-0.115\ep-0.031\ep^2-0.018\ep^3
\right ).
$$

iii) Assume that the estimate of Ref.\cite{BslPade} is accurate.
Expressed in terms of the pole mass the three-loop coefficient of the
 width is then a factor $0.7$ smaller than
the large $\beta_0$ approximation.
%%The exact result is then smaller by a factor $0.7$ than
%%the large $\beta_0$ approximation.
Rewriting the decay width through the $1S$ mass 
 and using $\bar \alpha_s^{(4)} = 0.22$ we obtain:
$$
\Gamma_{\rm sl} = \frac {G_F^2|V_{ub}|^2 {m_{1S}}^5}{192\pi^3} 
\left ( 
1-0.115\ep-0.031\ep^2 +0.008\ep^3
\right ).
$$

iv) One can see that at the two-loop level the exact result
  for the decay width (in terms of the pole mass) is about 
a factor $0.8$ smaller than the large $\beta_0$ approximation. 
Let us imagine that in the three-loop order the exact result is about 
half the size of the large $\beta_0$ approximation. We note that this 
%implies
 translates into
$\Delta \approx 80$ in Eq.(\ref{rate0}). 
 For $\bar \alpha_s^{(4)} = 0.22$ we obtain:
$$
\Gamma_{\rm sl} = \frac {G_F^2|V_{ub}|^2 {m_{1S}}^5}{192\pi^3} 
\left ( 
1-0.115\ep-0.031\ep^2+0.035\ep^3
\right ).
$$

We see therefore that there is a window for the parameter $\Delta$
where a
%where different approximations to the full result work fairly
%well and the 
parameterization of the decay width through the 
$1S$ mass seems to succeed in making the perturbative series convergent.
However, there is one  unpleasant feature.
The magnitude of the ${\cal O}(\ep^3)$ term in the $\Upsilon$ expansion 
strongly  depends on the value of the strong coupling constant 
and this is because various terms which scale differently in $\alpha_s$ 
contribute to a given  term in the $\Upsilon$ 
expansion\footnote{In present example, the highest power
of $\alpha_s$ in the $\ep^3$ terms is six and this contribution comes
from the leading order difference between the $1S$ and the pole mass.}.
In general, this dependence is quite strong. For example, by using 
$\bar \alpha_s=0.25$ we would have obtained $0.06$ as the coefficient
of the $\ep^3$ term in the fourth example. 

\section{Conclusions}
%{\it Conclusions.}
The main result of this letter, an analytic three-loop relation  
between the $\overline {\rm MS}$ and the pole masses, is given 
by Eq.(\ref{main}).  It  provides a first step towards a more 
 precise analysis of QCD effects in different low-scale shortx-distance 
 processes.  

Although this result is in good 
agreement with the large $\beta_0$ approximation, we have 
argued that it is difficult to  understand the structure 
of subleading terms.
% and for this reason the success of this approximation seems accidental. 
Nevertheless, our analysis of semileptonic $B$ decay $B \to X_u e \nu$
shows that it is likely 
  that the general physical picture 
of improving a badly behaved perturbative series at low orders
 by eliminating  the pole quark mass 
 remains valid also in the three-loop order.
 However, to answer this question unambiguously, one would have to perform 
  a three-loop QCD calculation of semileptonic decays of a heavy quark and 
 this seems to be a very non-trivial task at present. 
 It may be that the large $N_c$ approximation will
provide enough technical simplifications to be feasible beyond the
presently known perturbative orders 
and at the same time be accurate enough to be meaningful.

\section*{Acknowledgments}

We are grateful to M.~Beneke, D.~Broadhurst, J.H.~K\"uhn, S.~Larin 
 and  N.~Uraltsev for useful comments. 
 This research was supported in part by the United States
Department of Energy, contract DE-AC03-76SF00515, 
by BMBF under grant number BMBF-057KA92P, by
Gra\-duier\-ten\-kolleg ``Teil\-chen\-phy\-sik'' at the University of
Karlsruhe and by the DFG Forschergruppe 
``Quantenfeldtheorie, Computeralgebra und Monte-Carlo-Simulation''. 

%\bibliographystyle{../../pro/tex/prsty}
%\bibliography{../../pro/tex/phd}

\begin{thebibliography}{10}

\bibitem{msscheme} G. 't Hooft, Nucl. Phys.  {\bf B61}, 455 (1973).

\bibitem{bardeen} W.A.~Bardeen, A.J.~Buras D.W.~Duke and T.~Muta,
Phys. Rev.{\bf D18}, 3998 (1978). 

\bibitem{thooft} G.~'t~Hooft and M.~Veltman, Nucl. Phys. {\bf B44}, 189 (1972).


\bibitem{delphi}
P.~Abreu {\it et al.}, Phys. Lett.{\bf B418}, 431 (1998).

\bibitem{tarrach} R.~Tarrach, Nucl. Phys. {\bf B183}, 384 (1981).

\bibitem{broadhurst}
N. Gray,  D.J.~Broadhurst, W.~Grafe and K.~Schilcher,
 Z. Phys. {\bf C48}, 673 (1990).
 D.J.~Broadhurst, N.~Gray and K.~Schilcher,
Z. Phys. {\bf C52}, 111 (1991).

\bibitem{fleischer} J.~Fleischer, F.~Jegerlehner,~O.V.~Tarasov
and O.I.~Veretin, Nucl. Phys. {\bf B539}, 671 (1999).

\bibitem{kronfeld} A.S.~Kronfeld, Phys. Rev.{\bf D58}, (1998)
051501.

\bibitem{bigi} I.I~Bigi and N.G.~Uraltsev, Phys. Lett.{\bf B321},
412 (1994); I.I.~Bigi {\it et al.}, Phys. Rev.{\bf D50}, 2234 (1994).

\bibitem{braun} M.~Beneke and V.M.~Braun, Nucl. Phys.{\bf B426}, 301
(1994).

\bibitem{kinetic} I.I.~Bigi {\em et al.}, Phys. Rev. {\bf D56},
4017 (1996).

\bibitem{1s} A.H.~Hoang and T.~Teubner, hep-ph/9904468.

\bibitem{ps} M.~Beneke, Phys. Lett. {\bf B434}, 115 (1998).

\bibitem{hoang} A.H.~Hoang, hep-ph/9905550.

\bibitem{my}  K.~Melnikov and A.~Yelkhovsky, Phys. Rev. {\bf D59} (1999), 
114009.

\bibitem{beneke} M.~Beneke and A.~Signer, hep-ph/9906475.

\bibitem{bbb0} P.~Ball, M.~Beneke and V.M.~Braun, 
Nucl. Phys. {\bf B452}, 563 (1995).

\bibitem{chst}
K.~G.~Chetyrkin and M.~Steinhauser,
Phys.Rev.Lett. {\bf 83}, 4001 (1999); hep-ph/9911434.

\bibitem{andim3} O.V.~Tarasov, 
preprint JINR P2-82-900 (Dubna, 1982), unpublished;\\
S.A.~Larin, hep-ph/9302240; In Proceedings of the International School
``Particles and Cosmology'', April 1993, Baksan Neutrino Observatory of INR,
Russia, edt. by E.N.~Alekseev {\it et al.} (World Scientific, 1994).

%\bibitem{4looprenorm}
%        S.A. Larin, T. van Ritbergen and J.A.M. Vermaseren,
%          Phys. Lett. {\bf B400} 379 (1997);
%          Phys. Lett.  {\bf B405} 327 (1997);
%          K.G. Chetykin, Phys. Lett. {\bf B404}
%            161 (1997). 

\bibitem{dcpl1} B. Ovrut and H. Schnitzer,
             Nucl.Phys. {\bf B189}, 509 (1981);\\
        W. Bernreuther and W. Wetzel,
              Nucl. Phys. {\bf B197}, 228 (1982);\\
         W. Bernreuther,  Annals of Physics {\bf 151}, 127 (1983).

\bibitem{dcpl2} S.A. Larin, T. van Ritbergen and J.A.M. Vermaseren,
          Nucl. Phys. {\bf B438}, 278 (1995);
         K.G. Chetyrkin, B.A. Kniehl and  M. Steinhauser,
         Phys. Rev. Lett. {\bf 79 }, 2184 (1997).
    
    
\bibitem{ibp}  F.~Tkachov, Phys.Lett. {\bf B100}, 65 (1981);
     K.G.~Chetyrkin and F.~Tkachov, Nucl. Phys. {\bf B192},
 159 (1981).

\bibitem{Laporta} S. Laporta and E. Remiddi, Phys. Lett. {\bf B379}, 
          283 (1996).

\bibitem{schroeder} M. Peter, Phys. Rev. Lett. {\bf 78}, 602 (1997);\\
               Y.~Schr\"oder, Phys. Lett. {\bf B447}, 321 (1998).

\bibitem{largenexp} G. 't Hooft, Nucl. Phys. {\bf B72}, 461 (1974).

\bibitem{npb} I.I.~Bigi, N.G.~Uraltsev and A.I.~Vainshtein, 
Phys. Lett. {\bf B293}, 430 (1992); I.I.~Bigi {\it et al.}
Phys. Rev. Lett. {\bf 71}, 496 (1993).

\bibitem{timo} T. van Ritbergen, Phys.Lett. {\bf B454}, 353 (1999).

\bibitem{BslPade} M.R.~Ahmady, F.A. Chistie, V.~Elias and T.G.~Steele,
 hep-ph/9910551.

\bibitem{upexp} A.~Hoang, Z.~Ligeti and A.V.~Manohar, 
Phys. Rev. Lett. {\bf 82}, 277 (1999).

\bibitem{bbb} M.~Beneke and V.M.~Braun, Phys. Lett. {\bf B348}, 513 (1995);
P.~Ball, M.~Beneke  and V.M.~Braun, 
Phys.Rev. {\bf D52}, 3929 (1995).

\bibitem{bes}  I.~Bigi, M.~Shifman, N.~Uraltsev, A.~Vainshtein,
Phys.Rev. {\bf D56}, 4017 (1997).

\end{thebibliography}

\end{document}